\documentclass[traditabstract]{aa} 
\usepackage{epsfig,amsmath}
\usepackage[dvips]{color}
\usepackage{verbatim}
\usepackage{graphicx}
\usepackage{txfonts}
\usepackage{natbib}
\bibpunct{(}{)}{;}{a}{}{,}
\usepackage{hyperref}
\usepackage{rotating}
\usepackage{multicols}

\newcommand{\lya}{Ly$\alpha$}
\newcommand{\oi}{[O~{\sc i}]}

\newcommand{\hi}{H~{\sc i}}

\newcommand{\ciii}{C~{\sc iii}]}
\newcommand{\civ}{C~{\sc iv}}

\newcommand{\siii}{Si~{\sc ii}}

\newcommand{\siiv}{Si~{\sc iv}}
\definecolor{Black}{named}{Black}
\definecolor{Red}{named}{Red}
\definecolor{Green}{named}{Green}
\definecolor{Blue}{named}{Blue}

\begin{document}

\title{Compact to extended Lyman-$\alpha$ emitters in MAGPI: strong blue peak emission at $z\gtrsim3$} \titlerunning{Strong blue peak LAEs up to $z\sim4.8$}

\author{T. Mukherjee\inst{1}\fnmsep\inst{2}\fnmsep\inst{9}
          \and
          T. Zafar\inst{1}\fnmsep\inst{2}\fnmsep\inst{9}
          \and
          T. Nanayakkara\inst{3}\fnmsep\inst{9}
          \and
          E. Wisnioski \inst{4}\fnmsep\inst{9}
          \and
          A. Battisti \inst{4}\fnmsep\inst{9}
          \and
           A. Gupta \inst{5}\fnmsep\inst{9}
          \and
          C. D. P. Lagos \inst{6}\fnmsep\inst{9}
          \and
          K. E. Harborne \inst{6}\fnmsep\inst{9}
          \and
          C. Foster \inst{7}\fnmsep\inst{9}
          \and
          T. Mendel \inst{4}\fnmsep\inst{9}
          \and
          S. M. Croom \inst{8}\fnmsep\inst{9}
          \and
          A. Mailvaganam \inst{1}\fnmsep\inst{2}\fnmsep\inst{9}
          \and
          J. Prathap \inst{1}\fnmsep\inst{2}
          }
   \institute{School of Mathematical and Physical Sciences, Macquarie University, NSW 2109, Australia\\
   \email{tamal.mukherjee@hdr.mq.edu.au}
         \and
            Macquarie University Astrophysics and Space Technologies Research Centre, Sydney, NSW 2109, Australia
        \and 
           Centre for Astrophysics and Supercomputing, Swinburne University of Technology, P.O. Box 218, Hawthorn, 3122, VIC, Australia
        \and
            Research School of Astronomy and Astrophysics, Australian National University, Canberra, ACT 2611, Australia 
        \and
            International Centre for Radio Astronomy Research (ICRAR), Curtin University, Bentley WA, Australia
        \and
            International Centre for Radio Astronomy Research (ICRAR), The University of Western Australia, Crawley, WA 6009, Australia
        \and
            School of Physics, University of New South Wales, Sydney, NSW 2052, Australia
        \and
            Sydney Institute for Astronomy, School of Physics, University of Sydney, NSW 2006, Australia
        \and
            ARC Centre of Excellence for All Sky Astrophysics in 3 Dimensions (ASTRO-3D), Australia
        }
   \date{Received XX, XXXX; accepted XX, XXXX}
  \abstract
  {We report the discovery of three double-peaked Lyman-$\alpha$ emitters (LAEs) exhibiting strong blue peak emission at 2.9 $\lesssim z \lesssim$ 4.8, in the VLT/MUSE data obtained as part of the Middle Ages Galaxy Properties with Integral Field Spectroscopy (MAGPI) survey. These strong blue peak systems provide a unique window into the scattering of \lya\ photons by neutral hydrogen (\hi), suggesting gas inflows along the line-of-sight and low \hi\ column density. Two of them at $z=2.9$ and $z=3.6$ are spatially extended halos with their core regions clearly exhibiting stronger blue peak emissions than the red peak. However, spatial variations in the peak ratio and peak separation are evident over $25\times 26$\,kpc ($z=2.9$) and $19\times28$\,kpc ($z=3.6$) regions in these extended halos. 
  Notably, these systems do not fall in the regime of \lya\ blobs or nebulae. To the best of our knowledge, such a \lya\ halo with a dominant blue core has not been observed previously. 
  In contrast, the LAE at $z\sim4.8$ is a compact system spanning a $9\times9$\,kpc region and stands as the highest-redshift strong blue peak emitter ever detected. The peak separation of the bright cores in these three systems ranges from $\Delta_{\mathrm{peak}}\sim370$ to $660$\,km/s. The observed overall trend of decreasing peak separation with increasing radius is supposed to be controlled by \hi\ column density and gas covering fraction. Based on various estimations, in contrast to the compact LAE, our halos are found to be good candidates for LyC leakers. 
  These findings shed light on the complex interplay between \lya\ emission, gas kinematics, and ionising radiation properties, offering valuable insights into the evolution and nature of high-redshift galaxies.}
   \maketitle
%
\section{Introduction}
Investigation of the circum-galactic medium (CGM) plays a significant role in understanding the mechanisms driving galaxy formation and evolution. Interactions involving gas exchanges between galaxies and the surrounding intergalactic environment shape the evolution of galaxies. Lyman-$\alpha$ (Ly$\alpha$) emission is a key tracer of CGM gas around high-redshift star-forming galaxies. Extended \lya\ emission has been observed in various galactic environments: $(i)$ \lya\ halos around individual galaxies \citep{Kunth03, Steidel_2011, Matsuda_12, Hayes_2013, Leclercq_17}; $(ii)$ \lya\ blobs (LABs) associated with multiple sources \citep{Francis_96, Fynbo_99, Matsuda_11, Cai_2017}. The origin of this diffuse \lya\ emission in these halos remains a subject of ongoing debate, with several possibilities proposed, including scattering from star-forming regions, \lya\ fluorescence, gravitational cooling radiation from accretion or emission from satellite galaxies \citep{Haiman_2000,Zheng_2011, Byrohl_21}. 

\begin{figure*}
        \centering
        {\includegraphics[width=5.8cm,height=4.5cm,clip]{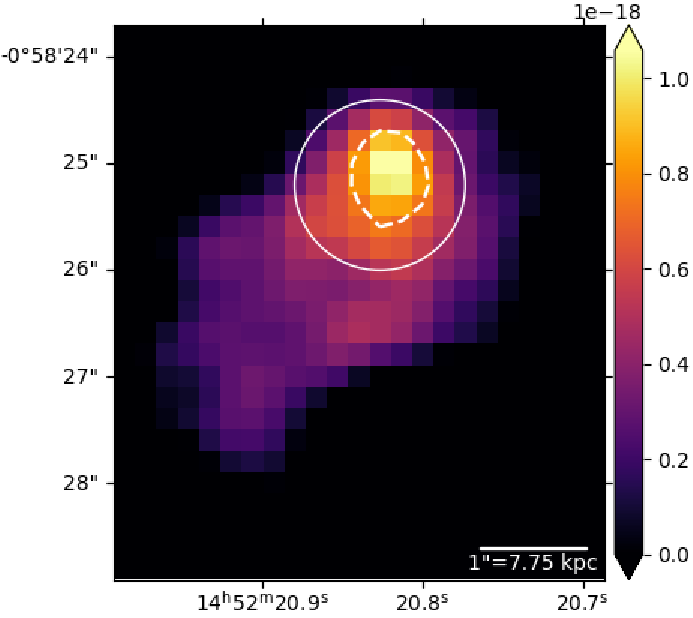}}
        {\includegraphics[width=5.8cm,height=4.5cm]{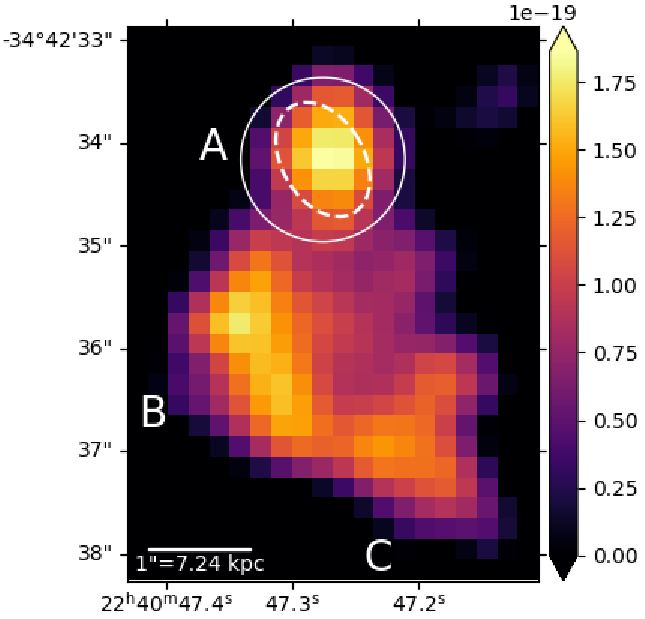}}
        {\includegraphics[width=5.8cm,height=4.5cm,clip]{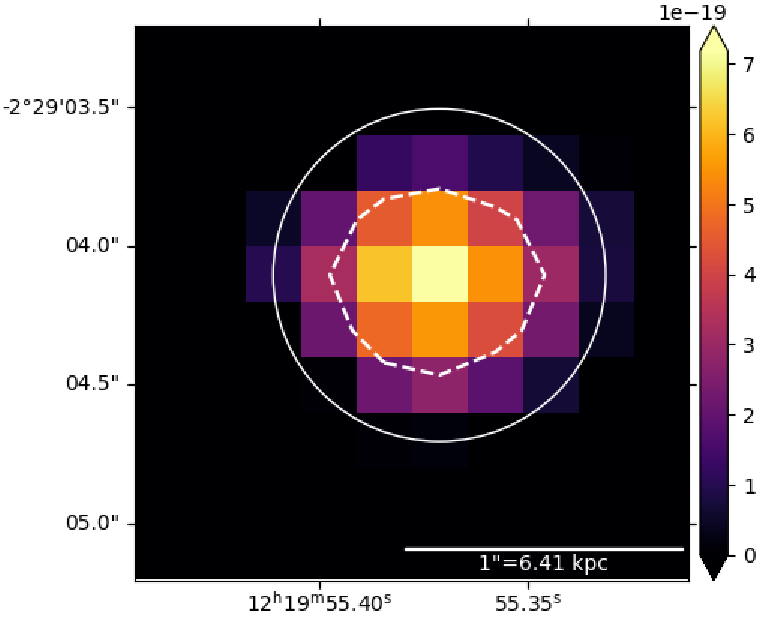}}
    \medskip
        {\includegraphics[width=5.8cm,height=4.5cm]{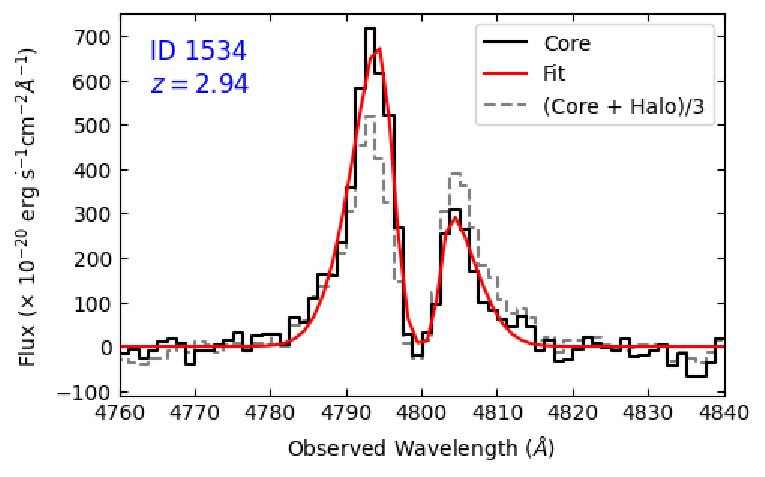}}
        {\includegraphics[width=5.8cm,height=4.5cm]{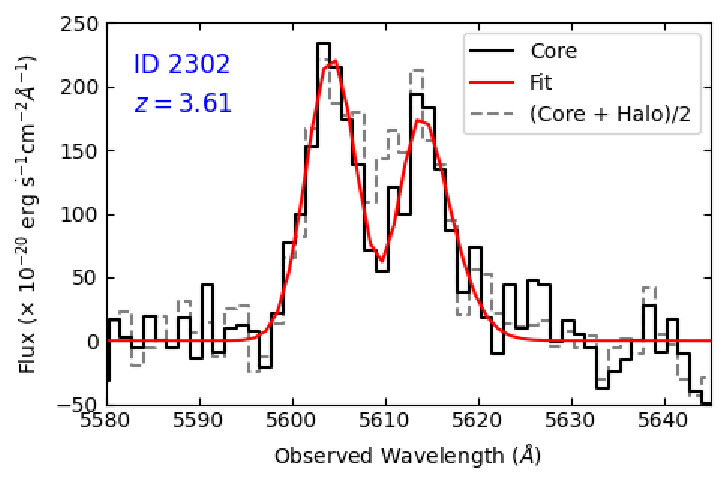}}
        {\includegraphics[width=5.8cm,height=4.5cm]{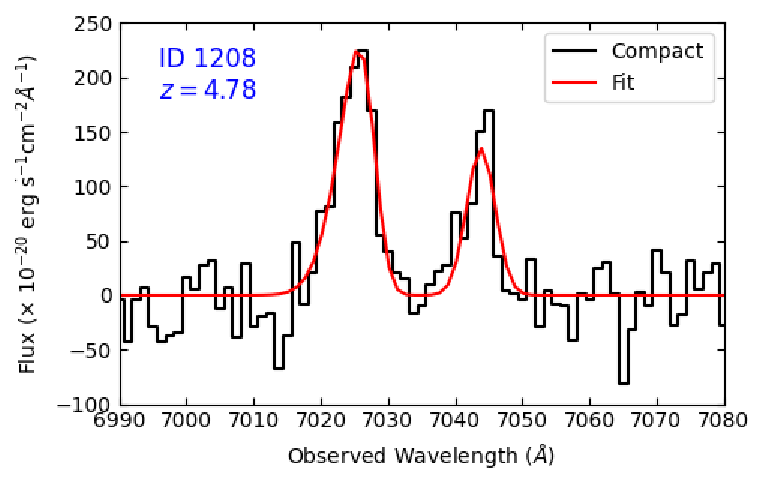}}
        \vspace{-0.4cm}\caption{Shown are continuum-subtracted \lya\ NB images smoothed with 2D Gaussian Kernel of $\sigma =1$ ({\bf top row}) and 1D \lya\ emission profiles fitted with a double-skewed Gaussian ({\bf bottom row}). The 1D spectra of the cores for $z=2.9$, $3.6$, and $4.8$ sources are extracted using $0.8''$, $0.8''$, \& $0.6''$ apertures (solid white circles in the NB images), respectively. These are shown in black solid lines in the bottom row. Dashed grey lines for $z=2.9$ and $z=3.6$ sources show the {\it scaled} \lya\ spectra of entire halos. While $z=4.8$ LAE, which is a compact source, is covered in its entirety within the $0.6''$ aperture. White dashed contours in the NBs indicate the position of the continuum. For the $z=3.6$ LAE, three distinct regions are labeled as A, B, \& C in the NB image. The colorbars are in the units of $\mathrm{erg}\,\mathrm{s}^{-1} \mathrm{cm}^{-2}\,\mathrm{pix}^{-2}$ where 1\,pix = $0.2''$.}
    \label{FigGam1}%
\end{figure*}

Complex radiative transfer and resonant scattering of \lya\ photons have given rise to diverse spectral profiles, with the double-peaked profile being particularly intriguing \citep{Gronke16,Verhamme17}. The escape of ionising photons ($>$13.6\,eV) into the intergalactic medium (IGM) played a pivotal role in cosmic reionisation, with faint, low-mass, early star-forming galaxies \citep{Bunker10,Finkelstein15} contributing significantly. However, the interplay between the emission and escape of ionising radiation, particularly Lyman-continuum (LyC) photons ($\lambda<912$\,\AA ), remains a puzzle. This enigma is partly due to the inherent opacity of the neutral IGM \citep{Madau95,Inoue14}. This opacity of the IGM to LyC radiation beyond $z\gtrsim4.5$ makes direct observation of LyC emission challenging \citep{Vanzella18}. Double-peaked profiles serve as a probe to indirectly infer characteristics related to LyC leakage at higher redshifts. The peak separation, correlated with the neutral hydrogen (\hi) column density ($N_{\mathrm{HI}}$), plays a pivotal role in determining the LyC escape fraction, $f^{\mathrm{LyC}}_{\mathrm{esc}}$, in high-redshift galaxies \citep{Verhamme15, Kakiichi21}. Furthermore, observations and simulations show a relation between $f^{\mathrm{LyC}}_{\mathrm{esc}}$ and \lya\ equivalent widths \citep[EWs;][]{Verhamme17,Maji22}. 

The majority of double-peaked profiles observed in \lya\ emitters \citep[LAEs;][]{Shapley03,Izotov18,Kerutt22} and LABs \citep[e.g.,][]{Matsuda06} predominantly exhibit a dominant red peak, indicating gas outflows. A dominant red peak is obvious as intergalactic absorption tends to extinguish the blue peak. However, some $z>6$ LAEs display double-peaked \lya\ profiles where the blue peak has a high probability of being scattered by the increasingly neutral IGM at higher redshifts \citep{Hu16, Hayes21, Endsley22}. Despite the common association of strong LAEs with low-redshift LyC leakers \citep{Bian20}, the opposite is not necessarily true. An interesting distinction arises: \lya\ double-peaked profiles with a stronger and broader blue peak than the red peak, implying gas inflows \citep{Blaizot23}, have been detected in two \lya\ nebulae at $z\sim 3$ \citep{Vanzella17}, two LABs \citep{Ao20,Li22} and, only in the outskirts of $z\sim 2$ \lya\ halos \citep{Erb18, Erb23}. Besides these extended halos and luminous blobs, where two peaks typically originate from different portions within an extended halo or spatially separated regions within the blob, only two compact and stronger blue peak LAEs have been reported to date \citep{Furtak22, Mc22}, claiming that both blue and red-peaks originate from the LAE itself. 

In this letter, we report the discovery of three strong ``blue peak'' LAEs, spanning across redshifts 2.9 $\lesssim z \lesssim$ 4.8 obtained as part of the Middle Ages Galaxy Properties with Integral Field Spectroscopy (MAGPI\footnote{Based on observations obtained at the Very Large Telescope (VLT) of the European Southern Observatory (ESO), Paranal, Chile (ESO program ID 1104.B-0526)}) survey \citep{Foster21}. The structure of this letter is as follows. In \S\,\ref{sec:obs}, we provide details of observations and data reduction including MAGPI survey \lya\ identification. In \S\,\ref{sec:results}, we present our results, followed by a discussion in \S\,\ref{sec:discuss}. Throughout this letter, we assume a standard flat $\Lambda$CDM cosmology with parameters $H_0$= 70 $\mathrm{km} \, \mathrm{s}^{-1} \mathrm{Mpc}^{-1}$, $\Omega_{\mathrm{m}}$ = 0.3 and  $\Omega_{\Lambda}$ = 0.7.

\section{Observations and data reduction}\label{sec:obs}
The MAGPI survey is an ongoing Large Program (Program ID: 1104.B-0526) on the VLT/MUSE, targeting 56 fields (with stellar masses $>$7$\times10^{10}$\,$M_\odot$). As of September 2023, 48 fields have been completed. The primary objective of this survey is to conduct a detailed spatially resolved spectroscopic analysis of stars and ionised gas within 0.25< $z$ <0.35 galaxies \citep[see][]{Foster21}. Data are taken using the MUSE Wide Field Mode ($1'\times1'$) with a spatial sampling rate of $0.2''$/pixel and an average image quality of $0.65''$\,FWHM. Each field is observed in six observing blocks, each comprising 2$\times$1320\,s exposures, resulting in a total integration time of 4.4\,hrs. The survey primarily employs the nominal mode, providing a wavelength coverage ranging from 4700\,\AA\ to 9350\,\AA, with a dispersion of 1.25\,\AA. Ground-Layer Adaptive Optics is used to correct atmospheric seeing effects, resulting in a 270\,\AA\ wide gap between 5780\,\AA\ and 6050\,\AA\ due to the GALACSI laser notch filter. Deeper MAGPI data resulted in detection of both foreground sources within the local universe and distant background sources, including LAEs up to $z\sim6.4$. 

\begin{figure*}
        \centering
        {\includegraphics[width=5.7cm,height=4.0cm,clip]{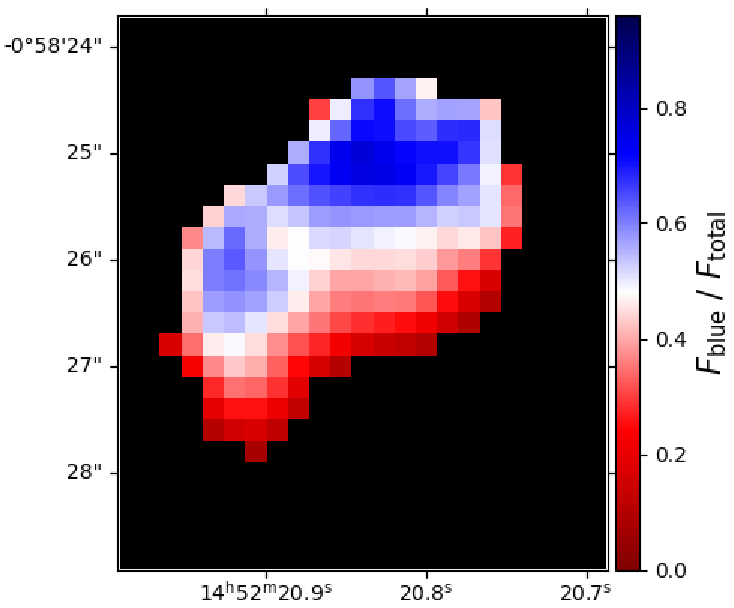}}
        {\includegraphics[width=5.7cm,height=4.0cm]{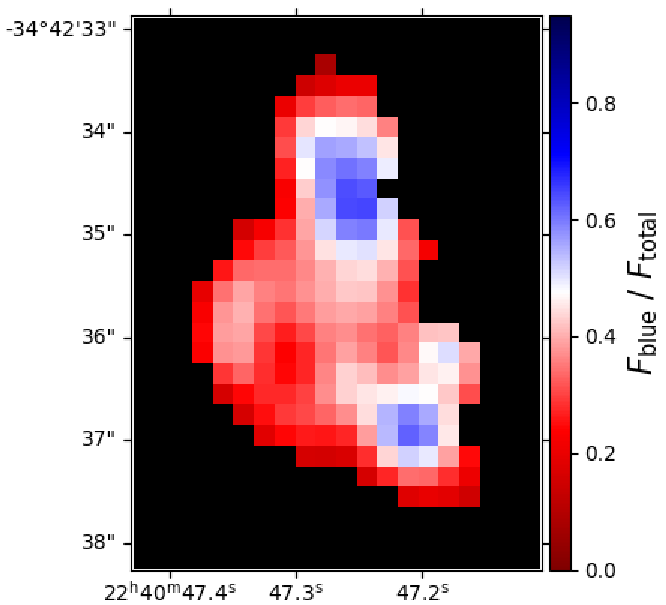}}
        {\includegraphics[width=5.7cm,height=4.0cm,clip]{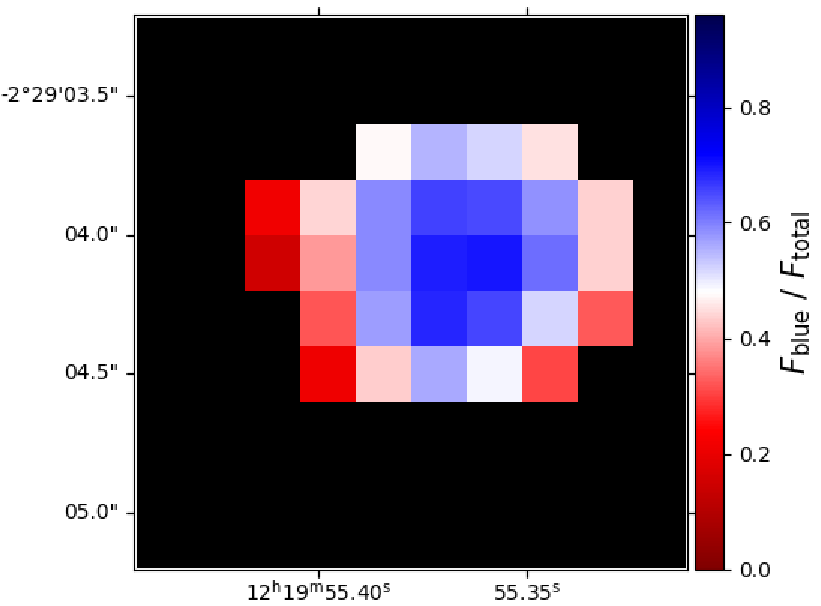}}
    \medskip
        {\includegraphics[width=5.7cm,height=4.0cm]{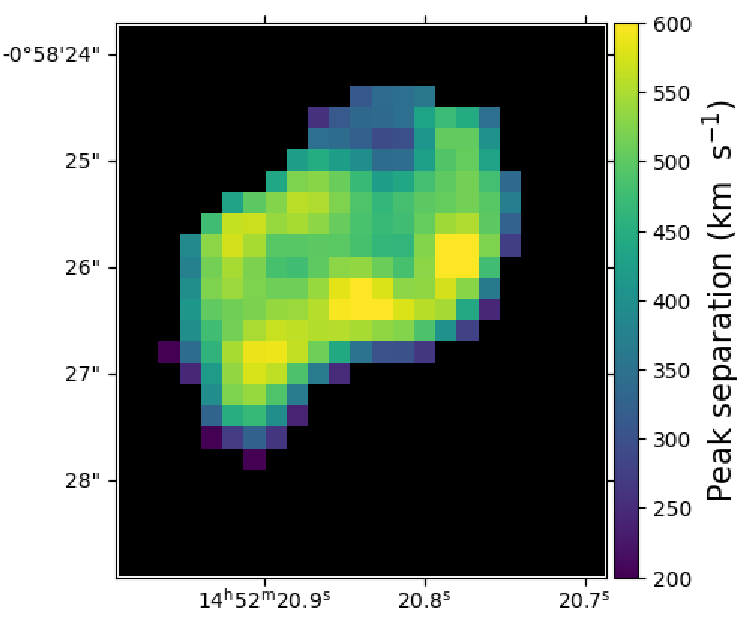}}
        {\includegraphics[width=5.7cm,height=4.0cm]{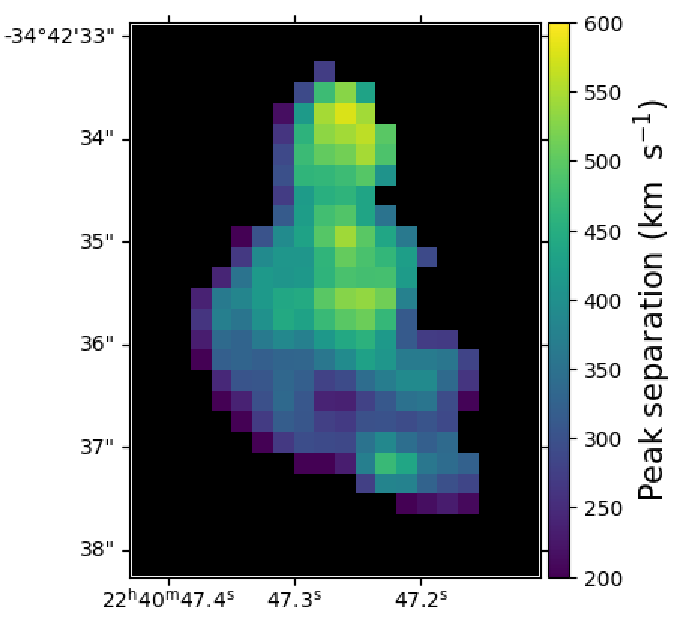}}
        {\includegraphics[width=5.7cm,height=4.0cm]{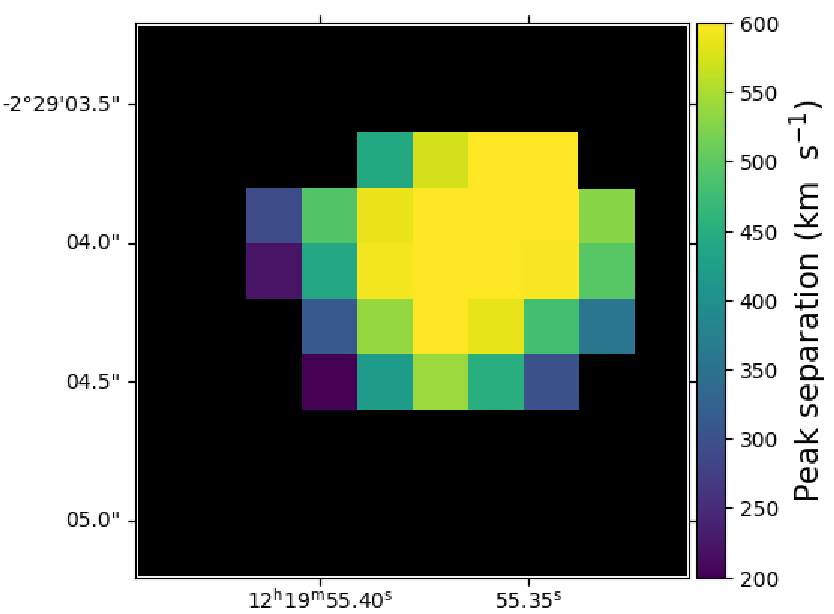}}
        \vspace{-0.3cm}\caption{{\bf Top row:} Pixel by pixel map color-coded by blue-to-total flux ratio; {\bf Bottom row:} Peak separation map (arranged from left to right in increasing order of redshift ($z=2.9$, $3.6$, and $4.8$) scaled within the range of $200 - 600$ km/s.}
    \label{FigGam2}%
\end{figure*}

\begin{table*}
\begin{minipage}[t]{\columnwidth}
\caption{Properties of strong blue peak LAEs.}      
\label{log_table} 
\centering
\renewcommand{\footnoterule}{}  
\setlength{\tabcolsep}{4.5pt}
\vspace{-0.2cm}
\begin{tabular}{l c c c c c c c c c c}\hline\hline
Field & RA & Dec & $z$ & Flux ratio & $\Delta_{\mathrm{peak}}$ & $f^{\mathrm{LyC}}_{\mathrm{esc}}$ & log $L$ & $\mathrm{EW}_0$ & $\mathrm{FWHM}_{\mathrm{blue}}$ & $\mathrm{FWHM}_{\mathrm{red}}$ \\
ID & & & & $F_{\mathrm{blue}}/F_{\mathrm{total}}$ & (km/s) &  & (ergs/s) & (\AA) & (km/s) & (km/s)\\
\hline
1534 & 14:52:20.88 & $-00$:58:25.20 & 2.94 & 0.74 & 370.14 & 0.050 & 42.65 & $60.67 \pm 5.43$ & $654.1 \pm 34.1$& $592.9 \pm 63.1$\\
2302 & 22:40:47.20 & $-34$:42:34.34 & 3.61 & 0.60 & 402.40 & 0.035 & 42.52 & $85.05\pm 25.0$ & $360.8 \pm 32.9$  & $446.4 \pm 84.7$ \\
1208 & 12:19:55.30 & $-02$:29:04.30 & 4.78 & 0.69 & 658.21 & 0.010 & 42.72 & $7.160 \pm 0.65$  & $415.9 \pm 61.7$  & $174.6 \pm 32.6$\\\hline
\end{tabular}
\end{minipage}
\end{table*}

MUSE data reduction were done using \texttt{pymusepipe2}, a Python wrapper for the MUSE reduction pipeline \citep{Weilbacher20}. A comprehensive discussion of the data reduction procedures will be presented in the first data release (Mendel et al. in prep.). \texttt{LSDCat} \citep{HW17} was used for faint sources particularly LAEs identification, accompanied by both automated and visual inspection. This extensive search led to the detection of 360 new LAEs distributed across 35 MAGPI fields (Mukherjee et al. in prep). Among these, a preliminary examination of LAE profiles revealed the presence of 35 double-peaked profiles out of which three distinct double-peak LAEs show strong blue peak emissions that are presented in this letter.

\section{Results}\label{sec:results}

We find dominant blue peak emission features in three newly discovered double-peak LAEs at redshifts $z=2.9$, $z=3.6$, and $z=4.8$, corresponding to MAGPI field IDs\,1534, 2302, and 1208, respectively (see Table\,\ref{log_table}). Alongside strong \lya\ emissions, the MUSE data also reveal various faint rest-frame UV emission and absorption lines: $i)$ For LAE at $z=2.9$, we identify emission lines such as \civ\ $\lambda 1548, 1550$ and \ciii\ $\lambda 1908$, as well as multiple absorption lines from both high (\siiv\ and \civ) and low ionisations (\siii\ and \oi) transitions. $ii)$ In the case of $z=3.6$ LAE, we tentatively observe \civ\ $\lambda 1548, 1550$ and \ciii\ $\lambda 1908$ emissions and absorption lines from \civ\ and low ionisation \siii\ transitions. $iii)$ For $z=4.8$ LAE, we detect \siiv\ $\lambda 1393, 1402$, and \civ\ $\lambda 1548, 1550$ emission lines, as well as absorption lines from high (\siiv\ and \civ) and low ionisation (\siii\ and \oi) transitions. For the former two, \ciii\ $\lambda 1908$ lines are either too faint or have low S/N to derive an accurate systemic redshift \citep{Bacon21} while for the latter, \ciii\ is outside the spectral window. Following \citet{Verhamme18}, an approximate systemic redshift can be derived by taking the average of the mean wavelengths of the two asymmetric Gaussians fitted to the two \lya\ peaks, which gives the following values: $z_{\mathrm{sys}} \sim 2.9484$, $z_{\mathrm{sys}} \sim 3.6136$ and $z_{\mathrm{sys}} \sim 4.788$.

For our detections, we use \texttt{Python}/\texttt{MPDAF} package \citep{Bacon16} designed for MUSE data cube analysis. We generate continuum-subtracted \lya\ narrow-band (NB) images from the MUSE data cube. We perform double-skewed Gaussian fitting to 1D \lya\ emission profiles, utilising the \texttt{Python}/\texttt{pyplatefit} package \citep{Bacon23}. \lya\ NB images along with corresponding 1D profiles are shown in Fig.\,\ref{FigGam1}. To quantify the strength of the blue peak, we measure blue-to-total flux ratios ($F_{\mathrm{blue}}/F_{\mathrm{total}}$; where $F_{\mathrm{total}} = F_{\mathrm{blue}} + F_{\mathrm{red}}$). Hence, $F_{\mathrm{blue}}/F_{\mathrm{total}} > 0.5$ implies a stronger blue peak. We also estimate other parameters including peak separations ($\Delta_{\mathrm{peak}}$), \lya\ luminosities, total rest-frame equivalent widths ($\mathrm{EW}_{0}$) and FWHMs of the blue and red-peaks. We determine the LyC escape fraction ($f^{\mathrm{LyC}}_{\mathrm{esc}}$) based on the empirical relationship between $f^{\mathrm{LyC}}_{\mathrm{esc}}$ and $\Delta_{\mathrm{peak}}$ from \citet{Izotov_Worseck_18}. These results, along with respective source coordinates (RA \& DEC of the bright cores), are summarised in Table\,\ref{log_table}. We also estimate LyC escape using \civ\ $\lambda 1550$/\ciii\ $\lambda 1908$ following the work of \citep{Schaerer22}. To investigate spatial variations in 1D profiles of our sources, we generate pixel-by-pixel maps color-coded by $F_{\mathrm{blue}}/F_{\mathrm{total}}$ and $\Delta_{\mathrm{peak}}$ on the \lya\ NB images (see Fig.\,\ref{FigGam2}).

\subsection{Spatial and spectral profiles of blue peak LAEs}
Our double-peaked \lya\ spectral profiles are unique, showing prominent and stronger blue peaks than the red peaks. 
We observe a variety of spatial distributions of \lya\ emissions, where the continuum-subtracted \lya\ NB images reveal two extended \lya\ halos and one compact, point-like LAE (Fig.\,\ref{FigGam1}, top row). Arranged by increasing redshift, these LAEs at $z=2.9$, $z=3.6$, and $z=4.8$ span regions measuring $24.8\times26.4$\,kpc, $19.2\times28.2$\,kpc, and $8.7\times8.7$\,kpc, respectively. Notably, the extended halos at $z=2.9$ and $z=3.6$ exhibit spatially varying spectra, featuring varying $F_{\mathrm{blue}}/F_{\mathrm{total}}$ and $\Delta_{\mathrm{peak}}$ in different spatial locations. In contrast, the compact LAE predominantly displays an overall blue peak dominated spectral profile, with mild variations in the outskirts. Below we break down our spectral and spatial analysis for each of the three cases:

$i)$ $z=2.9$ LAE (field ID\,1534) shows a compact `continuum-like' bright core and an extended halo (Fig.\,\ref{FigGam1}, left panel in top row). Such spatial profiles are characterised by a `two-component model' \citep[see][]{Wisotzki_16, Leclercq_17}. Two `tail-like' regions are identified within the halo. A faint continuum is also detected at the source location. Due to source crowding, faint continuum emissions from neighboring sources were also detected. To address this, we subtract the continuum emissions from neighboring sources to obtain pure emission lines of our source. Subsequently, we spectrally collapse this processed data cube over \lya\ wavelength range, to construct a continuum-subtracted \lya\ NB image showing an extensive \lya\ halo spanning $24.8\times26.4$ kpc. The halo displays spatially varying double-peaked spectra, with the brightest core featuring the strongest blue peak with $F_{\mathrm{blue}}/F_{\mathrm{total}}$ $= 0.74$ ($\mathrm{EW}_{\mathrm{blue,0}}$ $=$ 46.3$\pm$3.7\,\AA\ and $\mathrm{EW}_{\mathrm{red,0}}$ $=$ 14.4$\pm$1.7\,\AA).
Fig.\,\ref{FigGam1} (bottom row, left panel) presents a continuum-subtracted 1D profile of the core, extracted adopting a circular aperture of $0.8''$ radius. The overall 1D profile of the halo reveals a symmetric double peak with a slightly stronger blue peak. This spatial variation is evident in the $F_{\mathrm{blue}}/F_{\mathrm{total}}$ map (Fig.\,\ref{FigGam2}, top row, left panel), showing that the blue peak dominates in the bright core and its associated tail in the upper part of the halo while the red-peak dominates in the lower regions and in the outskirts. There is a smooth gradient from blue-dominated regions to red-dominated regions, with a boundary of $F_{\mathrm{blue}} = F_{\mathrm{red}}$ in the central region of the halo. We also observe a significant spatial variation in \lya\ peak separation (see Fig.\,\ref{FigGam2}, bottom row, left panel), ranging from $\sim 200$ to $750$\,km/s with larger separations ($\gtrsim 480$\,km/s) in the central region and smaller ($\lesssim 400$\,km/s) in the bright core and outskirts. We measure a \civ/\ciii\ ratio of 0.83 for this source.

$ii)$ We find an extended halo around the $z = 3.6$ LAE (field ID\,2302), showing three distinct bright regions labeled as A, B \& C (Fig.\,\ref{FigGam1}, top row middle panel), all displaying double-peak emission. Remarkably, these bright regions are spatially separated, a unique distribution seen in \lya\ halos. Due to source faintness and low S/N, a continuum light was not immediately visible. However, after applying a median-filter to the data cube, faint continuum light at the brightest core's location is revealed. Region\,A, the brightest core, displays prominent, strong blue peak emission with $F_{\mathrm{blue}}/F_{\mathrm{total}}$ $= 0.60$ ($\mathrm{EW}_{\mathrm{blue,0}}$ $=$ 47.2$\pm$13.0\,\AA\ and $\mathrm{EW}_{\mathrm{red,0}}$ $=$ 37.8$\pm$11.8\,\AA) 
and is encircled with an aperture of $0.8''$ radius. The entire halo shows a symmetric double-peaked profile with almost equal peak strengths.  
The flux-ratio map illustrates (see Fig.\,\ref{FigGam2}, top row middle panel) the blue component's dominance mainly in the bright core (A) and in the lower diffuse halo (C), with the red peak becoming more prominent as we move away from the core and other bright regions. We further observe larger peak separations ($\sim$400$-$600\,km/s) in region\,A and at the center of the halo, and regions B \& C exhibit peak separations of $\sim 390$ km/s, while comparatively lower peak separation ($\lesssim$370\,km/s) is seen in other regions within the halo and in the outskirts (Fig.\,\ref{FigGam2}, bottom row, middle panel). We find \civ/\ciii\ ratio to be 1.1.

$iii)$ The LAE at $z=4.8$ (field ID\,1208) shows a compact spatial profile (Fig.\,\ref{FigGam1}, top row, right panel) resembling `point-source LAEs' as studied by \cite{Bacon15}. Continuum light is detected at the location of \lya\ emission in this source. The \lya\ profile shows a broader and stronger blue peak emission. The flux-ratio map (Fig.\,\ref{FigGam2}, top row, right panel) confirms the prevalence of the blue component across most of the \lya\ emitting region, with slight red-peak dominance in the outskirts. Similar to previous cases, we find larger peak separations ($\gtrsim$600\,km/s) in the central region and smaller values ($\lesssim$450\,km/s) in the outskirts.

\section{Discussion}\label{sec:discuss}

In this letter, we report three new double-peaked LAEs with dominant blue peak emissions. These are $\sim 9$ \% of the sample of double-peaked LAEs found in the MAGPI survey, which is consistent with the results obtained in cosmological zoom-in simulations \citep{Blaizot23}. \lya\ emission profiles with dominant blue peaks have been previously observed in different spatially separated regions within LABs or luminous \lya\ nebulae. 
Comparing with the spatial size of LABs and nebulae \citep{Vanzella17,Li22}, we find that our extended LAEs are not in the regime of LABs or nebulae. However, we classify $z=2.9$ and $z=3.6$ LAEs as extended \lya\ halos \citep[see discussions in][]{Wisotzki_16,Leclercq_17}. The source at $z=2.9$ shows a similar spatial distribution as described by \citet{Wisotzki_16} and \citet{Leclercq_17}, featuring a continuum-like bright core surrounded by an extended halo. On the other hand, the extended source at $z=3.6$ reveals a more complex \lya\ spatial distribution, displaying three bright regions within the halo. This complexity suggests complex kinematics involving gas inflows and outflows in this galaxy \citep{Vanzella17, Ao20, Li22, Erb18}. This could also indicate a merger system, where two progenitors are merging into a single descendent \citep[e.g.,][]{Tilvi11}. A faint continuum light from these two sources is also detected in the MUSE data. However, lack of deep imaging data limits our ability to perform robust continuum subtraction and radial profile estimation. 

\citet{Erb18,Erb23} report complex kinematics in double-peaked extended \lya\ halos at $z \sim 2$, where they observe a dominant red peak in the central bright region with large peak separations 
, while a dominant blue peak at the edge of the halos where peak separations become narrower. 
For both of our extended halos, we find peak separation distributions similar to \citet{Erb18, Erb23}, but an inverse (core-to-halo) distributions for flux ratios. In both of our extended halos, the bright cores powering the \lya\ emission are dominated by blue peaks, while red peaks dominate as we move away from the bright cores towards the edges. This marks our discovery of the first two \lya\ halos at $z=2.9$ and $z=3.6$ with dominant blue cores, never observed before in \lya\ halos. 

In contrast, we find an overall blue peak dominated \lya\ spectral profile in the compact LAE at $z=4.8$. \citet{Furtak22} discover the first such compact lensed object at $z=3.2$, followed by a similar detection in a compact UV-bright star-forming galaxy at $z=3.6$ \citep{Mc22}. 
\citet{Furtak22} further find that both the blue and red peaks originate from the compact LAE itself. Similar to their flux ratio map, we observe that the entire \lya\ emitting region is dominated by the blue peak while the red peak becomes stronger at the edges. 

Overall, we observe that \lya\ peak separation decreases with increasing radius, consistent with \citet{Erb18,Erb23}. This trend is primarily driven by $N_{\mathrm{H\sc I}}$, where greater $N_{\mathrm{H\sc I}}$ values result in larger peak separations. \lya\ peak separation can be influenced by the cloud covering fraction (CF) within a clumpy outflow where higher CF mimicking higher $N_{\mathrm{H\sc I}}$ \citep{Gronke16} and impacting the velocity shift of escaping photons. However, this strongly depends on other parameters such as velocity and gas temperature \citep[e.g.][]{Kakiichi21,Li22}. 

\citet{Verhamme17} find that strong LyC emitters are strong LAEs with high rest-frame EWs ($\gtrsim 70$\,\AA) and large \lya\ escape fractions.
Simulations suggest a connection between double-peak separation and $N_{\mathrm{H\sc I}}$, indicating that for low $N_{\mathrm{H\sc I}}$, a narrow peak separation ($\lesssim 400$ km/s) serves as a strong indicator of LyC escape \citep{Verhamme15, Kakiichi21}. These predictions are confirmed observationally for low-redshift LyC leakers \citep{Verhamme17,Izotov21,Flury22}. Further, a high \civ/\ciii\ ratio ($\gtrsim 0.75$) has been proposed as a good tracer of LyC leakage in low-redshift galaxies \citep{Schaerer22}. We measure high rest-frame EWs in \lya, relatively narrow peak separations (see Table\,\ref{log_table}) and \civ/\ciii\ $> 0.75$ for the two extended LAEs, making them good candidates for LyC leakers. In contrast, we find a relatively large peak separation and a very low rest-frame EW for the compact LAE, implying a relatively high $N_{\mathrm{H\sc I}}$ in this system and almost no escape of LyC photons. 

Radiative transfer simulations, incorporating shells or clumpy, multi-phase models with varying outflow velocities typically focus on double-peaked profiles of \lya\ with stronger red peaks \citep{Gronke_Dijkstra_16, Erb23}. Cosmological hydrodynamics simulations suggest that red-dominated lines preferentially arise in face-on directions while blue-dominated lines are seen in the edge-on directions \citep{Blaizot23}. \lya\ line with a stronger blue peak than the red peak usually implies inflows of CGM gas along the line-of-sight during the accretion phase \citep{Vanzella17, Ao20, Blaizot23}. In this context, our new discoveries suggest inflowing gas systems or edge-on morphologies, warranting further investigation to understand complex gas kinematics and the environments of such rare LAEs to constrain the mechanisms that regulate their formation and evolution.

\begin{acknowledgements}
We wish to thank the ESO staff, and in particular the staff at Paranal Observatory, for carrying out the MAGPI observations. MAGPI targets were selected from GAMA. GAMA is a joint European-Australasian project based around a spectroscopic campaign using the Anglo-Australian Telescope. GAMA was funded by the STFC (UK), the ARC (Australia), the AAO, and the participating institutions. GAMA photometry is based on observations made with ESO Telescopes at the La Silla Paranal Observatory under programme ID 179.A-2004, ID 177.A-3016. The MAGPI team acknowledge support by the Australian Research Council Centre of Excellence for All Sky Astrophysics in 3 Dimensions (ASTRO 3D), through project number CE170100013.      
\end{acknowledgements}

\bibliographystyle{aa}
\bibliography{old_format.bib}{}

\begin{thebibliography}{53}
\expandafter\ifx\csname natexlab\endcsname\relax\def\natexlab#1{#1}\fi

\bibitem[{{Ao} {et~al.}(2020){Ao}, {Zheng}, {Henkel}, {Nie}, {Beelen}, {Cen},
  {Dijkstra}, {Francis}, {Geach}, {Kohno}, {Lehnert}, {Menten}, {Wang}, \&
  {Weiss}}]{Ao20}
{Ao}, Y., {Zheng}, Z., {Henkel}, C., {et~al.} 2020, Nature Astronomy, 4, 670

\bibitem[{{Bacon} {et~al.}(2023){Bacon}, {Brinchmann}, {Conseil}, {Maseda},
  {Nanayakkara}, {Wendt}, {Bacher}, {Mary}, {Weilbacher}, {Krajnovi{\'c}},
  {Boogaard}, {Bouch{\'e}}, {Contini}, {Epinat}, {Feltre}, {Guo}, {Herenz},
  {Kollatschny}, {Kusakabe}, {Leclercq}, {Michel-Dansac}, {Pello}, {Richard},
  {Roth}, {Salvignol}, {Schaye}, {Steinmetz}, {Tresse}, {Urrutia}, {Verhamme},
  {Vitte}, {Wisotzki}, \& {Zoutendijk}}]{Bacon23}
{Bacon}, R., {Brinchmann}, J., {Conseil}, S., {et~al.} 2023, \aap, 670, A4

\bibitem[{{Bacon} {et~al.}(2015){Bacon}, {Brinchmann}, {Richard}, {Contini},
  {Drake}, {Franx}, {Tacchella, S.}, {Vernet, J.}, {Wisotzki, L.}, {Blaizot,
  J.}, {Bouch\'e, N.}, {Bouwens, R.}, {Cantalupo, S.}, {Carollo, C. M.},
  {Carton, D.}, {Caruana, J.}, {Cl\'ement, B.}, {Dreizler, S.}, {Epinat, B.},
  {Guiderdoni, B.}, {Herenz, C.}, {Husser, T.-O.}, {Kamann, S.}, {Kerutt, J.},
  {Kollatschny, W.}, {Krajnovic, D.}, {Lilly, S.}, {Martinsson, T.},
  {Michel-Dansac, L.}, {Patricio, V.}, {Schaye, J.}, {Shirazi, M.}, {Soto, K.},
  {Soucail, G.}, {Steinmetz, M.}, {Urrutia, T.}, {Weilbacher, P.}, \& {de
  Zeeuw, T.}}]{Bacon15}
{Bacon}, R., {Brinchmann}, J., {Richard}, J., {et~al.} 2015, A\&A, 575, A75

\bibitem[{{Bacon} {et~al.}(2021){Bacon}, {Mary}, {Garel}, {Blaizot}, {Maseda},
  {Schaye}, {Wisotzki}, {Conseil}, {Brinchmann}, {Leclercq}, {Abril-Melgarejo},
  {Boogaard}, {Bouch{\'e}}, {Contini}, {Feltre}, {Guiderdoni}, {Herenz},
  {Kollatschny}, {Kusakabe}, {Matthee}, {Michel-Dansac}, {Nanayakkara},
  {Richard}, {Roth}, {Schmidt}, {Steinmetz}, {Tresse}, {Urrutia}, {Verhamme},
  {Weilbacher}, {Zabl}, \& {Zoutendijk}}]{Bacon21}
{Bacon}, R., {Mary}, D., {Garel}, T., {et~al.} 2021, \aap, 647, A107

\bibitem[{{Bacon} {et~al.}(2016){Bacon}, {Piqueras}, {Conseil}, {Richard}, \&
  {Shepherd}}]{Bacon16}
{Bacon}, R., {Piqueras}, L., {Conseil}, S., {Richard}, J., \& {Shepherd}, M.
  2016, {MPDAF: MUSE Python Data Analysis Framework}, Astrophysics Source Code
  Library, record ascl:1611.003

\bibitem[{{Bian} \& {Fan}(2020)}]{Bian20}
{Bian}, F. \& {Fan}, X. 2020, \mnras, 493, L65

\bibitem[{{Blaizot} {et~al.}(2023){Blaizot}, {Garel}, {Verhamme}, {Katz},
  {Kimm}, {Michel-Dansac}, {Mitchell}, {Rosdahl}, \& {Trebitsch}}]{Blaizot23}
{Blaizot}, J., {Garel}, T., {Verhamme}, A., {et~al.} 2023, \mnras, 523, 3749

\bibitem[{{Bunker} {et~al.}(2010){Bunker}, {Wilkins}, {Ellis}, {Stark},
  {Lorenzoni}, {Chiu}, {Lacy}, {Jarvis}, \& {Hickey}}]{Bunker10}
{Bunker}, A.~J., {Wilkins}, S., {Ellis}, R.~S., {et~al.} 2010, \mnras, 409, 855

\bibitem[{Byrohl {et~al.}(2021)Byrohl, Nelson, Behrens, Kostyuk, Glatzle,
  Pillepich, Hernquist, Marinacci, \& Vogelsberger}]{Byrohl_21}
Byrohl, C., Nelson, D., Behrens, C., {et~al.} 2021, \mnras, 506, 5129

\bibitem[{Cai {et~al.}(2017)Cai, Fan, Yang, Bian, Prochaska, Zabludoff,
  McGreer, Zheng, Green, Cantalupo, Frye, Hamden, Jiang, Kashikawa, \&
  Wang}]{Cai_2017}
Cai, Z., Fan, X., Yang, Y., {et~al.} 2017, \apj, 837, 71

\bibitem[{{Endsley} {et~al.}(2022){Endsley}, {Stark}, {Bouwens}, {Schouws},
  {Smit}, {Stefanon}, {Inami}, {Bowler}, {Oesch}, {Gonzalez}, {Aravena}, {da
  Cunha}, {Dayal}, {Ferrara}, {Graziani}, {Nanayakkara}, {Pallottini},
  {Schneider}, {Sommovigo}, {Topping}, {van der Werf}, \& {Hutter}}]{Endsley22}
{Endsley}, R., {Stark}, D.~P., {Bouwens}, R.~J., {et~al.} 2022, \mnras, 517,
  5642

\bibitem[{{Erb} {et~al.}(2023){Erb}, {Li}, {Steidel}, {Chen}, {Gronke},
  {Strom}, {Trainor}, \& {Rudie}}]{Erb23}
{Erb}, D.~K., {Li}, Z., {Steidel}, C.~C., {et~al.} 2023, \apj, 953, 118

\bibitem[{{Erb} {et~al.}(2018){Erb}, {Steidel}, \& {Chen}}]{Erb18}
{Erb}, D.~K., {Steidel}, C.~C., \& {Chen}, Y. 2018, \apjl, 862, L10

\bibitem[{{Finkelstein} {et~al.}(2015){Finkelstein}, {Ryan}, {Papovich},
  {Dickinson}, {Song}, {Somerville}, {Ferguson}, {Salmon}, {Giavalisco},
  {Koekemoer}, {Ashby}, {Behroozi}, {Castellano}, {Dunlop}, {Faber}, {Fazio},
  {Fontana}, {Grogin}, {Hathi}, {Jaacks}, {Kocevski}, {Livermore}, {McLure},
  {Merlin}, {Mobasher}, {Newman}, {Rafelski}, {Tilvi}, \&
  {Willner}}]{Finkelstein15}
{Finkelstein}, S.~L., {Ryan}, Russell~E., J., {Papovich}, C., {et~al.} 2015,
  \apj, 810, 71

\bibitem[{{Flury} {et~al.}(2022){Flury}, {Jaskot}, {Ferguson}, {Worseck},
  {Makan}, {Chisholm}, {Saldana-Lopez}, {Schaerer}, {McCandliss}, {Xu}, {Wang},
  {Oey}, {Ford}, {Heckman}, {Ji}, {Giavalisco}, {Amor{\'\i}n}, {Atek},
  {Blaizot}, {Borthakur}, {Carr}, {Castellano}, {De Barros}, {Dickinson},
  {Finkelstein}, {Fleming}, {Fontanot}, {Garel}, {Grazian}, {Hayes}, {Henry},
  {Mauerhofer}, {Micheva}, {Ostlin}, {Papovich}, {Pentericci}, {Ravindranath},
  {Rosdahl}, {Rutkowski}, {Santini}, {Scarlata}, {Teplitz}, {Thuan},
  {Trebitsch}, {Vanzella}, \& {Verhamme}}]{Flury22}
{Flury}, S.~R., {Jaskot}, A.~E., {Ferguson}, H.~C., {et~al.} 2022, \apj, 930,
  126

\bibitem[{{Foster} {et~al.}(2021){Foster}, {Mendel}, {Lagos}, {Wisnioski},
  {Yuan}, {D'Eugenio}, {Barone}, {Harborne}, {Vaughan}, {Schulze}, {Remus},
  {Gupta}, {Collacchioni}, {Khim}, {Taylor}, {Bassett}, {Croom}, {McDermid},
  {Poci}, {Battisti}, {Bland-Hawthorn}, {Bellstedt}, {Colless}, {Davies},
  {Derkenne}, {Driver}, {Ferr{\'e}-Mateu}, {Fisher}, {Gjergo}, {Johnston},
  {Khalid}, {Kobayashi}, {Oh}, {Peng}, {Robotham}, {Sharda}, {Sweet}, {Taylor},
  {Tran}, {Trayford}, {van de Sande}, {Yi}, \& {Zanisi}}]{Foster21}
{Foster}, C., {Mendel}, J.~T., {Lagos}, C.~D.~P., {et~al.} 2021, \pasa, 38,
  e031

\bibitem[{{Francis} {et~al.}(1996){Francis}, {Woodgate}, {Warren}, {M{\o}ller},
  {Mazzolini}, {Bunker}, {Lowenthal}, {Williams}, {Minezaki}, {Kobayashi}, \&
  {Yoshii}}]{Francis_96}
{Francis}, P.~J., {Woodgate}, B.~E., {Warren}, S.~J., {et~al.} 1996, \apj, 457,
  490

\bibitem[{{Furtak} {et~al.}(2022){Furtak}, {Plat}, {Zitrin}, {Topping},
  {Stark}, {Strait}, {Charlot}, {Coe}, {Andrade-Santos}, {Brada{\v{c}}},
  {Bradley}, {Lemaux}, \& {Sharon}}]{Furtak22}
{Furtak}, L.~J., {Plat}, A., {Zitrin}, A., {et~al.} 2022, \mnras, 516, 1373

\bibitem[{Fynbo {et~al.}(1999)Fynbo, Møller, \& Warren}]{Fynbo_99}
Fynbo, J.~U., Møller, P., \& Warren, S.~J. 1999, \mnras, 305, 849

\bibitem[{{Gronke} \& {Dijkstra}(2016)}]{Gronke_Dijkstra_16}
{Gronke}, M. \& {Dijkstra}, M. 2016, \apj, 826, 14

\bibitem[{{Gronke} {et~al.}(2016){Gronke}, {Dijkstra}, {McCourt}, \&
  {Oh}}]{Gronke16}
{Gronke}, M., {Dijkstra}, M., {McCourt}, M., \& {Oh}, S.~P. 2016, \apjl, 833,
  L26

\bibitem[{{Haiman} {et~al.}(2000){Haiman}, {Spaans}, \&
  {Quataert}}]{Haiman_2000}
{Haiman}, Z., {Spaans}, M., \& {Quataert}, E. 2000, \apjl, 537, L5

\bibitem[{Hayes {et~al.}(2013)Hayes, Östlin, Schaerer, Verhamme, Mas-Hesse,
  Adamo, Atek, Cannon, Duval, Guaita, Herenz, Kunth, Laursen, Melinder,
  Orlitová, Otí-Floranes, \& Sandberg}]{Hayes_2013}
Hayes, M., Östlin, G., Schaerer, D., {et~al.} 2013, \apjl, 765, L27

\bibitem[{{Hayes} {et~al.}(2021){Hayes}, {Runnholm}, {Gronke}, \&
  {Scarlata}}]{Hayes21}
{Hayes}, M.~J., {Runnholm}, A., {Gronke}, M., \& {Scarlata}, C. 2021, \apj,
  908, 36

\bibitem[{{Herenz} \& {Wisotzki}(2017)}]{HW17}
{Herenz}, E.~C. \& {Wisotzki}, L. 2017, \aap, 602, A111

\bibitem[{{Hu} {et~al.}(2016){Hu}, {Cowie}, {Songaila}, {Barger},
  {Rosenwasser}, \& {Wold}}]{Hu16}
{Hu}, E.~M., {Cowie}, L.~L., {Songaila}, A., {et~al.} 2016, \apjl, 825, L7

\bibitem[{{Inoue} {et~al.}(2014){Inoue}, {Shimizu}, {Iwata}, \&
  {Tanaka}}]{Inoue14}
{Inoue}, A.~K., {Shimizu}, I., {Iwata}, I., \& {Tanaka}, M. 2014, \mnras, 442,
  1805

\bibitem[{{Izotov} {et~al.}(2018{\natexlab{a}}){Izotov}, {Schaerer}, {Worseck},
  {Guseva}, {Thuan}, {Verhamme}, {Orlitov{\'a}}, \& {Fricke}}]{Izotov18}
{Izotov}, Y.~I., {Schaerer}, D., {Worseck}, G., {et~al.} 2018{\natexlab{a}},
  \mnras, 474, 4514

\bibitem[{{Izotov} {et~al.}(2021){Izotov}, {Worseck}, {Schaerer}, {Guseva},
  {Chisholm}, {Thuan}, {Fricke}, \& {Verhamme}}]{Izotov21}
{Izotov}, Y.~I., {Worseck}, G., {Schaerer}, D., {et~al.} 2021, \mnras, 503,
  1734

\bibitem[{{Izotov} {et~al.}(2018{\natexlab{b}}){Izotov}, {Worseck}, {Schaerer},
  {Guseva}, {Thuan}, {Fricke}, \& {Orlitov{\'a}}}]{Izotov_Worseck_18}
{Izotov}, Y.~I., {Worseck}, G., {Schaerer}, D., {et~al.} 2018{\natexlab{b}},
  \mnras, 478, 4851

\bibitem[{{Kakiichi} \& {Gronke}(2021)}]{Kakiichi21}
{Kakiichi}, K. \& {Gronke}, M. 2021, \apj, 908, 30

\bibitem[{{Kerutt} {et~al.}(2022){Kerutt}, {Wisotzki}, {Verhamme}, {Schmidt},
  {Leclercq}, {Herenz}, {Urrutia}, {Garel}, {Hashimoto}, {Maseda}, {Matthee},
  {Kusakabe}, {Schaye}, {Richard}, {Guiderdoni}, {Mauerhofer}, {Nanayakkara},
  \& {Vitte}}]{Kerutt22}
{Kerutt}, J., {Wisotzki}, L., {Verhamme}, A., {et~al.} 2022, \aap, 659, A183

\bibitem[{{Kunth} {et~al.}(2003){Kunth}, {Leitherer}, {Mas-Hesse},
  {{\"O}stlin}, \& {Petrosian}}]{Kunth03}
{Kunth}, D., {Leitherer}, C., {Mas-Hesse}, J.~M., {{\"O}stlin}, G., \&
  {Petrosian}, A. 2003, \apj, 597, 263

\bibitem[{{Leclercq} {et~al.}(2017){Leclercq}, {Bacon}, {Wisotzki}, {Mitchell},
  {Garel, Thibault}, {Verhamme, Anne}, {Blaizot, J\'er\'emy}, {Hashimoto,
  Takuya}, {Herenz, Edmund Christian}, {Conseil, Simon}, {Cantalupo,
  Sebastiano}, {Inami, Hanae}, {Contini, Thierry}, {Richard, Johan}, {Maseda,
  Michael}, {Schaye, Joop}, {Marino, Raffaella Anna}, {Akhlaghi, Mohammad},
  {Brinchmann, Jarle}, \& {Carollo, Marcella}}]{Leclercq_17}
{Leclercq}, F., {Bacon}, R., {Wisotzki}, L., {et~al.} 2017, A\&A, 608, A8

\bibitem[{{Li} {et~al.}(2022){Li}, {Steidel}, {Gronke}, {Chen}, \&
  {Matsuda}}]{Li22}
{Li}, Z., {Steidel}, C.~C., {Gronke}, M., {Chen}, Y., \& {Matsuda}, Y. 2022,
  \mnras, 513, 3414

\bibitem[{{Madau}(1995)}]{Madau95}
{Madau}, P. 1995, \apj, 441, 18

\bibitem[{{Maji} {et~al.}(2022){Maji}, {Verhamme}, {Rosdahl}, {Garel},
  {Blaizot}, {Mauerhofer}, {Pittavino}, {Victoria Feser}, {Chuniaud}, {Kimm},
  {Katz}, \& {Haehnelt}}]{Maji22}
{Maji}, M., {Verhamme}, A., {Rosdahl}, J., {et~al.} 2022, \aap, 663, A66

\bibitem[{{Marques-Chaves} {et~al.}(2022){Marques-Chaves}, {Schaerer},
  {{\'A}lvarez-M{\'a}rquez}, {Verhamme}, {Ceverino}, {Chisholm}, {Colina},
  {Dessauges-Zavadsky}, {P{\'e}rez-Fournon}, {Saldana-Lopez}, {Upadhyaya}, \&
  {Vanzella}}]{Mc22}
{Marques-Chaves}, R., {Schaerer}, D., {{\'A}lvarez-M{\'a}rquez}, J., {et~al.}
  2022, \mnras, 517, 2972

\bibitem[{{Matsuda} {et~al.}(2006){Matsuda}, {Yamada}, {Hayashino}, {Yamauchi},
  \& {Nakamura}}]{Matsuda06}
{Matsuda}, Y., {Yamada}, T., {Hayashino}, T., {Yamauchi}, R., \& {Nakamura}, Y.
  2006, \apjl, 640, L123

\bibitem[{Matsuda {et~al.}(2011)Matsuda, Yamada, Hayashino, Yamauchi, Nakamura,
  Morimoto, Ouchi, Ono, Kousai, Nakamura, Horie, Fujii, Umemura, \&
  Mori}]{Matsuda_11}
Matsuda, Y., Yamada, T., Hayashino, T., {et~al.} 2011, \mnras~Letters, 410, L13

\bibitem[{{Matsuda} {et~al.}(2012){Matsuda}, {Yamada}, {Hayashino}, {Yamauchi},
  {Nakamura}, {Morimoto}, {Ouchi}, {Ono}, {Umemura}, \& {Mori}}]{Matsuda_12}
{Matsuda}, Y., {Yamada}, T., {Hayashino}, T., {et~al.} 2012, \mnras, 425, 878

\bibitem[{{Schaerer} {et~al.}(2022){Schaerer}, {Izotov}, {Worseck}, {Berg},
  {Chisholm}, {Jaskot}, {Nakajima}, {Ravindranath}, {Thuan}, \&
  {Verhamme}}]{Schaerer22}
{Schaerer}, D., {Izotov}, Y.~I., {Worseck}, G., {et~al.} 2022, \aap, 658, L11

\bibitem[{{Shapley} {et~al.}(2003){Shapley}, {Steidel}, {Pettini}, \&
  {Adelberger}}]{Shapley03}
{Shapley}, A.~E., {Steidel}, C.~C., {Pettini}, M., \& {Adelberger}, K.~L. 2003,
  \apj, 588, 65

\bibitem[{Steidel {et~al.}(2011)Steidel, Bogosavljević, Shapley, Kollmeier,
  Reddy, Erb, \& Pettini}]{Steidel_2011}
Steidel, C.~C., Bogosavljević, M., Shapley, A.~E., {et~al.} 2011, \apj, 736,
  160

\bibitem[{{Tilvi} {et~al.}(2011){Tilvi}, {Scannapieco}, {Malhotra}, \&
  {Rhoads}}]{Tilvi11}
{Tilvi}, V., {Scannapieco}, E., {Malhotra}, S., \& {Rhoads}, J.~E. 2011,
  \mnras, 418, 2196

\bibitem[{{Vanzella} {et~al.}(2017){Vanzella}, {Balestra}, {Gronke}, {Karman},
  {Caminha}, {Dijkstra}, {Rosati}, {De Barros}, {Caputi}, {Grillo}, {Tozzi},
  {Meneghetti}, {Mercurio}, \& {Gilli}}]{Vanzella17}
{Vanzella}, E., {Balestra}, I., {Gronke}, M., {et~al.} 2017, \mnras, 465, 3803

\bibitem[{{Vanzella} {et~al.}(2018){Vanzella}, {Nonino}, {Cupani},
  {Castellano}, {Sani}, {Mignoli}, {Calura}, {Meneghetti}, {Gilli}, {Comastri},
  {Mercurio}, {Caminha}, {Caputi}, {Rosati}, {Grillo}, {Cristiani}, {Balestra},
  {Fontana}, \& {Giavalisco}}]{Vanzella18}
{Vanzella}, E., {Nonino}, M., {Cupani}, G., {et~al.} 2018, \mnras, 476, L15

\bibitem[{{Verhamme} {et~al.}(2018){Verhamme}, {Garel}, {Ventou}, {Contini},
  {Bouch{\'e}}, {Herenz}, {Richard}, {Bacon}, {Schmidt}, {Maseda}, {Marino},
  {Brinchmann}, {Cantalupo}, {Caruana}, {Cl{\'e}ment}, {Diener}, {Drake},
  {Hashimoto}, {Inami}, {Kerutt}, {Kollatschny}, {Leclercq}, {Patr{\'\i}cio},
  {Schaye}, {Wisotzki}, \& {Zabl}}]{Verhamme18}
{Verhamme}, A., {Garel}, T., {Ventou}, E., {et~al.} 2018, \mnras, 478, L60

\bibitem[{{Verhamme} {et~al.}(2015){Verhamme}, {Orlitov{\'a}}, {Schaerer}, \&
  {Hayes}}]{Verhamme15}
{Verhamme}, A., {Orlitov{\'a}}, I., {Schaerer}, D., \& {Hayes}, M. 2015, \aap,
  578, A7

\bibitem[{{Verhamme} {et~al.}(2017){Verhamme}, {Orlitov{\'a}}, {Schaerer},
  {Izotov}, {Worseck}, {Thuan}, \& {Guseva}}]{Verhamme17}
{Verhamme}, A., {Orlitov{\'a}}, I., {Schaerer}, D., {et~al.} 2017, \aap, 597,
  A13

\bibitem[{{Weilbacher} {et~al.}(2020){Weilbacher}, {Palsa}, {Streicher},
  {Bacon}, {Urrutia}, {Wisotzki}, {Conseil}, {Husemann}, {Jarno}, {Kelz},
  {P{\'e}contal-Rousset}, {Richard}, {Roth}, {Selman}, \&
  {Vernet}}]{Weilbacher20}
{Weilbacher}, P.~M., {Palsa}, R., {Streicher}, O., {et~al.} 2020, \aap, 641,
  A28

\bibitem[{{Wisotzki} {et~al.}(2016){Wisotzki}, {Bacon}, {Blaizot},
  {Brinchmann}, {Herenz}, {Schaye}, {Bouch\'e}, {Cantalupo}, {Contini, T.},
  {Carollo, C. M.}, {Caruana, J.}, {Courbot, J.-B.}, {Emsellem, E.}, {Kamann,
  S.}, {Kerutt, J.}, {Leclercq, F.}, {Lilly, S. J.}, {Patr\'{\i}cio, V.},
  {Sandin, C.}, {Steinmetz, M.}, {Straka, L. A.}, {Urrutia, T.}, {Verhamme,
  A.}, {Weilbacher, P. M.}, \& {Wendt, M.}}]{Wisotzki_16}
{Wisotzki}, L., {Bacon}, R., {Blaizot}, J., {et~al.} 2016, A\&A, 587, A98

\bibitem[{{Zheng} {et~al.}(2011){Zheng}, {Cen}, {Weinberg}, {Trac}, \&
  {Miralda-Escud{\'e}}}]{Zheng_2011}
{Zheng}, Z., {Cen}, R., {Weinberg}, D., {Trac}, H., \& {Miralda-Escud{\'e}}, J.
  2011, \apj, 739, 62

\end{thebibliography}


\end{document}